\documentclass[12pt]{article}
\usepackage{oldgerm}
\usepackage{yfonts}
\usepackage{euler}
\usepackage{graphics}
\usepackage{bm}
\usepackage{graphicx}
\usepackage{epstopdf}
\usepackage{amsmath}
\usepackage{amssymb}
\usepackage{amstext}
\usepackage{amscd}
\usepackage{amsfonts}
\usepackage{appendix}
\usepackage{color}

\DeclareMathAlphabet{\EuFrak}{U}{euf}{m}{n}
\DeclareMathAlphabet{\EuScript}{U}{eus}{m}{n}

\newcommand{\nd}{\noindent}

\newcommand{\be}{\begin{equation}}
\newcommand{\ee}{\end{equation}}
\newcommand{\ben}{\begin{eqnarray}}
\newcommand{\een}{\end{eqnarray}}

\title{{\bf New mathematics for the non
additive Tsallis' scenario}}

\author{{G. L. Ferri$^{1}$,F. Pennini$^{1,2}$, A. Plastino$^{3,5,6}$,
M.C.Rocca$^{3,4,5}$}\\
\small{$^{1}$Facultad de Ciencias Exactas y Naturales,}\\
\small{Universidad Nacional de La Pampa, Peru 151, 6300 Santa Rosa,}\\
\small{La Pampa, Argentina}\\
\small{Uruguay 151, Santa Rosa, La Pampa, Argentina}\\
\small{$^{2}$Universidad Cat\'olica del Norte, Av.~Angamos~0610, Antofagasta, Chile.}\\
\small{$^3$ Departamento de F\'{\i}sica,
Universidad Nacional de La Plata,}\\
\small{$^4$ Departamento de Matem\'{a}tica,
Universidad Nacional de La Plata,}\\
\small{$^5$ Consejo Nacional de Investigaciones Cient\'{\i}ficas
y Tecnol\'{o}gicas}\\
\small{(IFLP-CCT-CONICET)-C. C. 727, 1900 La Plata -
Argentina}\\\small{$^6$  SThAR - EPFL, Lausanne, Switzerland}}

\date{\today}

\begin{document}

\maketitle

\begin{abstract}

In this manuscript we investigate quantum uncertainties in a Tsallis' non additive scenario. To such an end
we appeal to q-exponentials, that are the  cornerstone of Tsallis' theory. In this respect, it is found that some new mathematics is needed and we are led to  construct a set of  novel special states that are the q-exponential equivalents of the ordinary coherent states of the harmonic oscillator. We then  characterize these new Tsallis' special states by
obtaining the associated i) probability distributions for a state of momentum $k$,  ii) mean values for some functions of space an momenta,  and iii) concomitant
quantum uncertainties. The latter are then compared to the usual ones.
\nd {\bf Keywords:} Tsallis' Statistics, quantum uncertainties, q-exponentials.

\end{abstract}

\newpage

\renewcommand{\theequation}{\arabic{section}.\arabic{equation}}

\section{Introduction}

\nd During more than 25 years, an important topic in statistical
mechanics theory revolved around the notion of generalized
non additive statistics, pioneered by Tsallis \cite{tsallis88}. It has been
amply demonstrated that, in many occasions,  the celebrated
Boltzmann-Gibbs logarithmic entropy does not yield a correct
description of the system under scrutiny \cite{tsallisbook}. Other
entropic forms, called non additive entropies $S_q$ ($q \in {\cal R}$), produce a much better
performance \cite{tsallisbook}. The non additive law reads, for two independent systems A and B,  
$S_q(AB)= S(A)+ S(B) + (1-q) S(A)S(B)$.   
One may cite a large number of
such instances. For example, non-ergodic systems exhibiting  a
complex dynamics \cite{tsallisbook}.

\nd The non-extensive statistical mechanics of Tsallis' has been
employed to fruitfully discuss phenomena in variegated fields. One
may mention, for instance, high-energy physics
\cite{[4]}-\cite{[44]}, spin-glasses \cite{[5]}, cold atoms in
optical lattices \cite{[6]}, trapped ions \cite{[7]}, anomalous
diffusion \cite{[8]}, \cite{[9]}, dusty plasmas \cite{[10]},
low-dimensional dissipative and conservative maps in  dynamical
systems \cite{[11]}, \cite{[12]}, \cite{[13]}, turbulent flows
\cite{[14]}, Levy flights \cite{wilk}, the QCD-based Nambu, Jona,
Lasinio  model of a many-body field theory \cite{15}, etc. Notions
related to q-statistical mechanics have been found useful not only
in physics but also in chemistry, biology, mathematics, economics,
informatics, and quantum mechanics  \cite{[17]}, \cite{[18]}, \cite{[19]},  \cite{nrt}.
 {\it Given the importance of the Tsallis-materials, the associated mathematics acquires particular relevance.
We believe to be here making some interesting contributions to such mathematics}.

\vskip 3mm
 \nd  The probability distribution associated to the non additive,  q-statistics is the so-called q-exponential \cite{tsallisbook},
that becomes the customary exponential (CE) in the limit $q
\rightarrow 1$. Physical states described via q-exponentials (qEs)
are the focus of our present concerns.  We obtain them by replacing
CEs by  (qEs) whenever physical states expressed in CE-terms emerge.
A reference to coherent states is then needed (see for instance
\cite{Carmichael}). Then, with regards to the line of inquire just
mentioned,  we construct the q-equivalents if coherent states which
are special forms of q-exponentials. We characterize the ensuing
q-equivalents by evaluation of its main properties, and then discuss
the  associated quantum uncertainties. A note of warning is due here. Our new q-equivalents {\it have 
 nothing to do with the so-called q-deformed coherent states of Quesne, Eremin-Meldianov, and others. These are coherent states of a deformed harmonic oscillator} \cite{quesne}. 
\setcounter{equation}{0}

\section{Prerequisites}

Let us briefly remind the reader of the coherent states of the harmonic
oscillator (HO) $|\alpha>$, or Glauber states
\cite{Gla63a,Gla63b,Gla63c}. A coherent state $|\alpha \rangle$ is a
specific kind of quantum state of minimum uncertainty, the one that most resembles a
classical state. It is applicable to the quantum harmonic
oscillator, the electromagnetic field, etc., and describes a maximal
kind of coherence and a classical kind of behavior.
 The states $ |\alpha\rangle$ are normalized, i.e.,
$\langle \alpha|\alpha\rangle=1$, and they provide us with a
resolution of the identity operator

\be \int \frac{\mathrm{d}^2 \alpha}{\pi}\,|\alpha\rangle\langle
\alpha|=1, \ee which is a completeness relation for the coherent
states~\cite{Gla63c}. The standard coherent states
$|\alpha\rangle$ for the harmonic oscillator are eigenstates  of
the annihilation operator $\hat a$, with complex eigenvalues

\be \label{alfa} \alpha=\frac {q + ip} {\sqrt{2}},\ee which
satisfy $\hat a |\alpha\rangle=\alpha
|\alpha\rangle$~\cite{Gla63c}.\vskip 3mm

\nd The $n-$th HO eigenfunction is
\begin{equation} \label{eq2.1}
\phi_n(x)=\left(\frac {m\omega} {\hbar}\right)^{ \frac {1}
{4}}{\cal H}_n\left(\sqrt{\frac {m\omega} {\hbar}}x\right),
\end{equation}
where ${\cal H}_n$ is Hermite's $n-$th order generalized function
\begin{equation}
\label{eq2.2} {\cal H}_n(x)=\left(\pi^{\frac {1} {2}}2^n
n!\right)^{- \frac {1} {2}} e^{-\frac {x^2} {2}} H_n(x),
\end{equation}
while $H_n$ is the concomitant Hermite polynomial. In the
x-representation, the coherent state reads
\begin{equation}
\label{eq2.3} \psi_{\alpha}(x)= e^{-\frac {|\alpha|^2} {2}}
\sum\limits_{n=0}^{\infty}\frac {\alpha^n} {\sqrt{n!}}\phi_n(x),
\end{equation}
or
\begin{equation}
\label{eq2.4} \psi_{\alpha}(x)=\left(\frac {m\omega}
{\hbar}\right)^{ \frac {1} {4}} e^{-\frac {|\alpha|^2} {2}}
\sum\limits_{n=0}^{\infty}\frac {\alpha^n} {\sqrt{n!}} {\cal
H}_n\left(\sqrt{\frac {m\omega} {\hbar}}x\right).
\end{equation}
For convenience we choose $\sqrt{\frac {m\omega} {\hbar}}=1$.
Thus, for the HO we have
\begin{equation}
\label{eq2.5} \phi_n(x)={\cal H}_n\left(x\right),
\end{equation}
and for its coherent states (CS)
\begin{equation}
\label{eq2.6} \psi_{\alpha}(x)= e^{-\frac {|\alpha|^2} {2}}
\sum\limits_{n=0}^{\infty}\frac {\alpha^n} {\sqrt{n!}} {\cal
H}_n\left(x\right).
\end{equation}
\nd
We  use at this point  the interesting fact that the CS can be made to compactly read (see Appendix A)
\begin{equation}
\label{eq2.7} \psi_{\alpha}(x)=\pi^{-\frac {1} {4}} e^{-\frac
{\alpha^2} {2}} e^{-\frac {|\alpha|^2} {2}} e^{-\frac {x^2} {2}}
e^{\sqrt{2}\alpha x}.
\end{equation}
To prove that (\ref{eq2.7}) is equal to (\ref{eq2.6})
we expand (\ref{eq2.7}) \`a la Hermite
\begin{equation}
\label{eq2.8} \psi_{\alpha}(x)=\sum\limits_{n=0}^{\infty} a_n{\cal
H}_n(x),
\end{equation}
and compute  $a_n$ as
\begin{equation}
\label{eq2.9} a_n=\int\limits_{-\infty}^{\infty}
\psi_{\alpha}(x){\cal H}_n(x)\;dx.
\end{equation}
Accordingly,
\begin{equation}
\label{eq2.10} a_n= \pi^{-\frac {1} {4}} e^{-\frac {\alpha^2} {2}}
e^{-\frac {|\alpha|^2} {2}} \int\limits_{-\infty}^{\infty}
e^{-\frac {x^2} {2}} e^{\sqrt{2}\alpha x} {\cal H}_n(x)\;dx,
\end{equation}
that can be recast as
\begin{equation}
\label{eq2.11} a_n= \frac {\pi^{-\frac {1} {4}} e^{-\frac
{|\alpha|^2} {2}}} {\left(n!2^n\pi^{\frac {1} {2}}\right)^{ \frac
{1} {2}}} \int\limits_{-\infty}^{\infty} e^{-\left(x-\frac
{\alpha} {\sqrt{2}}\right)^2} H_n(x)\;dx.
\end{equation}
We appeal now to an Integral-Table result (see \cite{tp1}) to
obtain
\begin{equation}
\label{eq2.12} a_n= \frac {\pi^{-\frac {1} {4}} e^{-\frac
{|\alpha|^2} {2}}} {\left(n!2^n\pi^{\frac {1} {2}}\right)^{ \frac
{1} {2}}} \pi^{\frac {1} {2}}2^{\frac {n} {2}}\alpha^n,
\end{equation}
or
\begin{equation}
\label{eq2.13} a_n=(n!)^{-\frac {1} {2}}\alpha^n e^{-\frac
{|\alpha|^2} {2}}.
\end{equation}
Replacing now (\ref{eq2.13}) into (\ref{eq2.8}) we reach
(\ref{eq2.6}) and prove  (\ref{eq2.7}).
Our results in this paper are based on the equation (\ref{eq2.7}),
translated into q-parlance.

\setcounter{equation}{0}

\section{Special  states associated to the non additive, q-statistics}
We start here work in this respect, and wish to report some
advances.
An extremely  important and critical  result is (\ref{eq2.7}) for an ordinary  coherent state, that we will q-generalize via replacement CE $\rightarrow$ qE.
 The ensuing state, that one may call a Tsallis' pseudo-coherent one,  is obtained, we reiterate,  by
replacing the exponential (\ref{eq2.7})  by the associated
q-exponential $e_q(x)$ \cite{tsallisbook}
 \be e_q(x) = [1+ (1-q)x]^{1/1-q}; \,\,q \in \mathcal{R}, \ee
that becomes the ordinary exponential at $q=1$. Accordingly, we
have

\begin{equation}
\label{eq3.1} \psi_{\alpha q}(x)=A(q,\alpha) \left[1+\frac {q-1}
{2}\left(x^2-2\sqrt{2}\alpha x+
|\alpha|^2+\alpha^2\right)\right]^{\frac {1} {1-q}},
\end{equation}
where $A(q,\alpha)$ is a normalization constant to be determined. 
Remember that these states {\it have 
 nothing to do with the so-called q-coherent states of Quesne, Eremin-Meldianov, and others} \cite{quesne}.

\vskip 3mm \nd  {\it We proceed now
to determine the mathematical apparatus
associated to these states $\psi_{\alpha q}$, i.e., 1) normalization,
2) overlaps, 3) probability distributions (PD) 4) mean values, and 5) uncertainties, in order to describe the nature of our special states, which is the goal of this paper}.
\vskip 3mm

\nd We need to appeal to some cumbersome mathematics. In particular, Lauricella functions $F_D$, described
in Appendix B, become of the essence. They are extensions to several variables of the hypergeometric functions.

\subsection{Normalization}

 For our present, new q-states we need, first
of all,  an explicit expression for the overlap involved in the
normalization process
\[<\psi_{\alpha q}|\psi_{\alpha q}>=A^2(q,\alpha)
\int\limits_{-\infty}^{\infty}
\left[1+\frac {q-1} {2}\left(x^2-2\sqrt{2}\alpha x+
|\alpha|^2+\alpha^2\right)\right]^{\frac {1} {1-q}}
\otimes\]
\begin{equation}
\label{eq3.2} \left[1+\frac {q-1}
{2}\left(x^2-2\sqrt{2}\alpha^{\ast} x+
|\alpha|^2+\alpha^{\ast2}\right)\right]^{\frac {1} {1-q}}dx.
\end{equation}
This necessitates appeal to Lauricella functions $F_D$.  We recast  (\ref{eq3.2}) in the form
\[<\psi_{\alpha q}|\psi_{\alpha q}>=A^2(q,\alpha)
\left(\frac {q-1} {2}\right)^{\frac {2} {1-q}}
\int\limits_{-\infty}^{\infty}
\left(x-\sqrt{2}\alpha -
\sqrt{|\alpha|^2+\alpha^2-\frac {2} {q-1}}
\right)^{\frac {1} {1-q}}\otimes\]
\[\left(x-\sqrt{2}\alpha +
\sqrt{|\alpha|^2+\alpha^2-\frac {2} {q-1}}
\right)^{\frac {1} {1-q}}\otimes\]
\[\left(x-\sqrt{2}\alpha^{\ast}-
\sqrt{|\alpha|^2+\alpha^{\ast 2}-\frac {2} {q-1}}
\right)^{\frac {1} {1-q}}
\otimes\]
\begin{equation}
\label{eq3.3} \left(x-\sqrt{2}\alpha^{\ast}+
\sqrt{|\alpha|^2+\alpha^{\ast 2}-\frac {2} {q-1}} \right)^{\frac
{1} {1-q}}dx.
\end{equation}
Utilizing Eq.  (\ref{eqa3}) from Appendix B we find
\[<\psi_{\alpha q}|\psi_{\alpha q}>=A^2(q,\alpha)
\frac {q-1} {5-q}
\left(\frac {q-1} {2}\right)^{\frac {2} {1-q}}\otimes\]
\[F_D\left(\frac {5-q} {q-1}; \frac {1} {q-1},
\frac {1} {q-1},\frac {1} {q-1},\frac {1} {q-1},
\frac {4} {q-1};\right.\]
\[1+\sqrt{2}\alpha +
\sqrt{\alpha^2-|\alpha|^2-\frac {2} {q-1}},\]
\[1+\sqrt{2}\alpha -
\sqrt{\alpha^2-|\alpha|^2-\frac {2} {q-1}},\]
\[1+\sqrt{2}\alpha^{\ast} +
\sqrt{\alpha^{\ast 2}-|\alpha|^2-\frac {2} {q-1}},\]
\begin{equation}
\label{eq3.4} \left.1+\sqrt{2}\alpha^{\ast} - \sqrt{\alpha^{\ast
2}-|\alpha|^2-\frac {2} {q-1}}\right).
\end{equation}
Now, because of the normalization requirement
\begin{equation}
\label{eq3.5} <\psi_{\alpha q}|\psi_{\alpha q}>=1,
\end{equation}\newpage
we get for  the constant $A(q,\alpha)$ the expression
\[A(q,\alpha)=\left[
\frac {q-1} {5-q}
\left(\frac {q-1} {2}\right)^{\frac {2} {1-q}}\right.\]
\[F_D\left(\frac {5-q} {q-1}; \frac {1} {q-1},
\frac {1} {q-1},\frac {1} {q-1},\frac {1} {q-1},
\frac {4} {q-1};\right.\]
\[1+\sqrt{2}\alpha +
\sqrt{\alpha^2-|\alpha|^2-\frac {2} {q-1}},\]
\[1+\sqrt{2}\alpha -
\sqrt{\alpha^2-|\alpha|^2-\frac {2} {q-1}},\]
\[1+\sqrt{2}\alpha^{\ast} +
\sqrt{\alpha^{\ast 2}-|\alpha|^2-\frac {2} {q-1}},\]
\begin{equation}
\label{eq3.6} \left.\left.1+\sqrt{2}\alpha^{\ast} -
\sqrt{\alpha^{\ast 2}-|\alpha|^2-\frac {2} {q-1}}\right)
\right]^{-\frac {1} {2}}.
\end{equation}

\subsection{Scalar product}
Usual coherent states are not orthogonal. Again we will appeal to Lauricella functions $F_D$ (Appendix B). 
Thus, we compute now the scalar product (overlap)  of two
arbitrary states $\psi_{\alpha q}$

\[<\psi_{\alpha q}|\psi_{\beta q}>=A(q,\alpha)
A(q,\beta)\frac {q-1} {5-q}
\left(\frac {q-1} {2}\right)^{\frac {2} {1-q}}\otimes\]
\[F_D\left(\frac {5-q} {q-1}; \frac {1} {q-1},
\frac {1} {q-1},\frac {1} {q-1},\frac {1} {q-1},
\frac {4} {q-1};\right.\]
\[1+\sqrt{2}\alpha +
\sqrt{\alpha^2-|\alpha|^2-\frac {2} {q-1}},\]
\[1+\sqrt{2}\alpha -
\sqrt{\alpha^2-|\alpha|^2-\frac {2} {q-1}},\]
\[1+\sqrt{2}\beta^{\ast} +
\sqrt{\beta^{\ast 2}-|\beta|^2-\frac {2} {q-1}},\]
\begin{equation}
\label{eq3.7} \left.1+\sqrt{2}\beta^{\ast} - \sqrt{\beta^{\ast
2}-|\beta|^2-\frac {2} {q-1}}\right).
\end{equation}
The non-normalized Tsallis' pseudo-coherent state
\begin{equation}
\label{eqs.1}
\phi_{\alpha q}(x)=\frac {\psi_{\alpha q}(x)}
{A(q,\alpha)}
\end{equation}
is a proper vector corresponding to the
proper value $\alpha$ of the operator $a_q$
given by:
\begin{equation}
\label{eqs.2}
a_q(x)f(x)=\frac {x} {\sqrt{2}} f(x) +
\frac {f^{1-q}(x)} {\sqrt{2}} \frac {df(x)} {dx}
\end{equation}
Note when $q=1$, $a_q$ is the usual annihilation
operator of the Harmonic Oscillator.

\subsection{Associated probability distribution (PD)}

We pass now to the PD associated to a Tsallis pseudo-coherent state.
We start by noting that
\begin{equation}
\label{eq3.8} |\alpha, q>=A(q,\alpha)
\int\limits_{-\infty}^{\infty} \left[1+\frac {q-1}
{2}\left(x^2-2\sqrt{2}\alpha x+
|\alpha|^2+\alpha^2\right)\right]^{\frac {1} {1-q}} |x>\;dx.
\end{equation}
Thus, the overlap between a plane wave of momentum $k$ and
$|\alpha, q>$ is
\begin{equation}
\label{eq3.9} <k|\alpha, q>=\frac {A(q,\alpha)} {\sqrt{2\pi}}
\int\limits_{-\infty}^{\infty} e^{-ikx} \left[1+\frac {q-1}
{2}\left(x^2-2\sqrt{2}\alpha x+
|\alpha|^2+\alpha^2\right)\right]^{\frac {1} {1-q}} dx,
\end{equation}
that can be rewritten as
\[<k|\alpha, q>=\frac {A(q,\alpha)} {\sqrt{2\pi}}
\int\limits_{-\infty}^{\infty}
e^{-ikx}
\left[x+\sqrt{2}\alpha+\sqrt{\alpha^2-
|\alpha|^2-\frac {2} {q-1}}\right]^{\frac {1} {1-q}}
\otimes\]
\begin{equation}
\label{eq3.10} \left[x+\sqrt{2}\alpha-\sqrt{\alpha^2-
|\alpha|^2-\frac {2} {q-1}}\right]^{\frac {1} {1-q}} dx.
\end{equation}
Using now the Integral-Table result  \cite{tp4} we find
\[<k|\alpha, q>=\frac {Sgn(k)\sqrt{2\pi} A(q,\alpha)
|k|^{\frac {3-q} {q-1}}} {\Gamma\left(\frac {2} {q-1}\right)}
e^{-\frac {i\pi Sgn(k)} {q-1}}\otimes\]
\begin{equation}
\label{eq3.11} e^{i\left(\sqrt{2}\alpha+\sqrt{\alpha^2-
|\alpha|^2-\frac {2} {q-1}}\right)} \phi\left(\frac {1} {q-1},
\frac {2} {q-1}; -2i\sqrt{\alpha^2-|\alpha|^2-\frac {2}
{q-1}}\;|k|\right).
\end{equation}
The  PD we are looking for becomes
\[|<k|\alpha, q>|^2=\frac {2\pi [A(q,\alpha)]^2
|k|^{\frac {6-2q} {q-1}}} {\left[\Gamma\left(\frac {2} {q-1}\right)\right]^2}\otimes\]
\[e^{i\left[\sqrt{2}(\alpha-\alpha^{\ast})+\sqrt{\alpha^2-
|\alpha|^2-\frac {2} {q-1}}-
\sqrt{\alpha^{\ast 2}-
|\alpha|^2-\frac {2} {q-1}}
\right]}\otimes\]

\[\phi\left(\frac {1} {q-1}, \frac {2} {q-1};
-2i\sqrt{\alpha^2-|\alpha|^2-\frac {2} {q-1}}\;|k|\right)\otimes\]
\begin{equation}
\label{eq3.12} \phi\left(\frac {1} {q-1}, \frac {2} {q-1};
2i\sqrt{\alpha^{\ast 2}-|\alpha|^2-\frac {2} {q-1}}\;|k|\right),
\end{equation}
and gives the probability of encountering momentum $k$ if the system
is described by $|\alpha, q>$.

\section{Towards determining uncertainties}

\nd We need to evaluate several mean values to this end.
\subsection{Mean value of $x^2$}

We can calculate now $<x^2>_q$. It is given by
\[<x^2>_q=A^2(q,\alpha)\left(\frac {q-1} {2}
\right)^{\frac {2} {1-q}}
\int\limits_{-\infty}^{\infty}
\left(x^2-2\sqrt{\alpha^{\ast}}x+|\alpha|^2+
\alpha^{\ast 2}+\frac {2} {q-1}\right)^{
\frac {1} {1-q}}
\otimes\]
\begin{equation}
\label{eq3.14}
x^2\left(x^2-2\sqrt{\alpha}x+|\alpha|^2+
\alpha^2+\frac {2} {q-1}\right)^{
\frac {1} {1-q}}dx
\end{equation}
Lets $\beta_1,\beta_2,\beta_3,\beta_4$ be given by
\[\beta_1=\sqrt{2}\alpha^{\ast}+
\sqrt{\alpha^{\ast 2}-|\alpha|^2-\frac {2} {q-1}}\]
\[\beta_2=\sqrt{2}\alpha^{\ast}-
\sqrt{\alpha^{\ast 2}-|\alpha|^2-\frac {2} {q-1}}\]
\[\beta_3=\sqrt{2}\alpha+
\sqrt{\alpha^2-|\alpha|^2-\frac {2} {q-1}}\]
\begin{equation}
\label{eq3.15}
\beta_4=\sqrt{2}\alpha-
\sqrt{\alpha^2-|\alpha|^2-\frac {2} {q-1}}
\end{equation}
Then we can write (\ref{eq3.14}) as
\[<x^2>_q=A^2(q,\alpha)\left(\frac {q-1} {2}\right)^{
\frac {2} {1-q}}
\int\limits_{-\infty}^{\infty}x^2
(x-\beta_1)^{\frac {1} {1-q}}
(x-\beta_2)^{\frac {1} {1-q}}
\otimes \]
\begin{equation}
\label{eq3.16}
(x-\beta_3)^{\frac {1} {1-q}}(x-\beta_4)^{\frac {1} {1-q}}\;dx
\end{equation}
Appealing now to (\ref{eqa3}), and to  Lauricella functions $F_D$,  we obtain for
(\ref{eq3.16})
\[<x^2>_q=2A^2(q,\alpha)\left(\frac {q-1} {2}\right)^{
\frac {2} {1-q}}\frac {\Gamma\left(\frac {7-3q} {q-1}\right)}
{\Gamma\left(\frac {4} {q-1}\right)}\]
\begin{equation}
\label{eq3.17}
F_D\left(\frac {7-3q} {q-1}; \frac {1} {q-1},\frac {1} {q-1},\frac {1} {q-1},
\frac {1} {q-1}; \frac {4} {q-1};1+\beta_1, 1+\beta_2,
1+\beta_3, 1+\beta_4\right)
\end{equation}

\subsection{Mean value of $x$}

Once again, we  appeal here to Lauricella functions $F_D$ (Appendix B). 
In the same way as above, we have for $<x>_q$ the expression
\[<x>_q=A^2(q,\alpha)\left(\frac {q-1} {2}\right)^{
\frac {2} {1-q}}\frac {\Gamma\left(\frac {6-2q} {q-1}\right)}
{\Gamma\left(\frac {4} {q-1}\right)}\]
\begin{equation}
\label{eq3.18}
F_D\left(\frac {6-2q} {q-1}; \frac {1} {q-1},\frac {1} {q-1},\frac {1} {q-1},
\frac {1} {q-1}; \frac {4} {q-1};1+\beta_1, 1+\beta_2,
1+\beta_3, 1+\beta_4\right)
\end{equation}

\subsection{Mean value of $p^2$}

The evaluation of $<p^2>_q$ is somewhat  more involved. For it, we
have
\[<p^2>_q=-A^2(q,\alpha)\left(\frac {q-1} {2}
\right)^{\frac {2} {1-q}}
\int\limits_{-\infty}^{\infty}
\left(x^2-2\sqrt{\alpha^{\ast}}x+|\alpha|^2+
\alpha^{\ast 2}+\frac {2} {q-1}\right)^{
\frac {1} {1-q}}
\otimes\]
\begin{equation}
\label{eq3.19}
\frac {\partial^2} {\partial x^2}
\left(x^2-2\sqrt{\alpha}x+|\alpha|^2+
\alpha^2+\frac {2} {q-1}\right)^{
\frac {1} {1-q}}dx
\end{equation}
or
\[<p^2>_q=-\frac {A^2(q,\alpha)} {(1-q)}
\left(\frac {q-1} {2}\right)^{\frac {2} {1-q}}
\left[2\int\limits_{-\infty}^{\infty}
\left(x^2-2\sqrt{\alpha^{\ast}}x+|\alpha|^2+
\alpha^{\ast 2}+\frac {2} {q-1}\right)^{
\frac {1} {1-q}}\right.\otimes\]
\[\left(x^2-2\sqrt{\alpha}x+|\alpha|^2+
\alpha^2+\frac {2} {q-1}\right)^{
\frac {q} {1-q}}dx\; + \]
\[\frac {q} {1-q}
\int\limits_{-\infty}^{\infty}
\left(x^2-2\sqrt{\alpha^{\ast}}x+|\alpha|^2+
\alpha^{\ast 2}+\frac {2} {q-1}\right)^{
\frac {1} {1-q}}(2x-2\sqrt{2}\alpha)^2\otimes\]
\begin{equation}
\label{eq3.20}
\left.\left(x^2-2\sqrt{\alpha}x+|\alpha|^2+
\alpha^2+\frac {2} {q-1}\right)^{
\frac {2q-1} {1-q}}dx\right]
\end{equation}
Appealing again to (\ref{eqa3}) the result for $<p^2>_q$ is
\[<p^2>_q=-\frac {A^2(q,\alpha)} {(1-q)}
\left(\frac {q-1} {2}\right)^{\frac {2} {1-q}}
\left\{2 \frac {\Gamma\left(\frac {3+q} {q-1}\right)}
{\Gamma\left(\frac {2q+2} {q-1}\right)}\right.
\otimes\]
\[F_D\left(\frac {3+q} {q-1}; \frac {1} {q-1},\frac {1} {q-1},\frac {q} {q-1},
\frac {q} {q-1}; \frac {2q+2} {q-1};1+\beta_1, 1+\beta_2,
1+\beta_3, 1+\beta_4\right)+\]
\[\frac {q} {1-q}\left[8
\frac {\Gamma\left(\frac {3+q} {q-1}\right)}
{\Gamma\left(\frac {4q} {q-1}\right)}\right.\otimes\]
\[F_D\left(\frac {3+q} {q-1}; \frac {1} {q-1},\frac {1} {q-1},
\frac {2q-1} {q-1},
\frac {2q-1} {q-1}; \frac {4q} {q-1};1+\beta_1, 1+\beta_2,
1+\beta_3, 1+\beta_4\right)-\]
\[8\sqrt{2}\alpha
\frac {\Gamma\left(\frac {2q+2} {q-1}\right)}
{\Gamma\left(\frac {4q} {q-1}\right)}\otimes\]
\[F_D\left(\frac {2q+2} {q-1}; \frac {1} {q-1},\frac {1} {q-1},
\frac {2q-1} {q-1},
\frac {2q-1} {q-1}; \frac {4q} {q-1};1+\beta_1, 1+\beta_2,
1+\beta_3, 1+\beta_4\right)+\]
\[8\alpha^2
\frac {\Gamma\left(\frac {3q+1} {q-1}\right)}
{\Gamma\left(\frac {4q} {q-1}\right)}\otimes\]
\begin{equation}
\label{eq3.21}
\left.\left.
F_D\left(\frac {3q+1} {q-1}; \frac {1} {q-1},\frac {1} {q-1},
\frac {2q-1} {q-1},
\frac {2q-1} {q-1}; \frac {4q} {q-1};1+\beta_1, 1+\beta_2,
1+\beta_3, 1+\beta_4\right)\right]\right\}
\end{equation}

\subsection{Mean value of $p$}

Analogously,  we have for $<p>_q$
\[<p>_q=-iA^2(q,\alpha)
\left(\frac {q-1} {2}
\right)^{\frac {2} {1-q}}
\int\limits_{-\infty}^{\infty}
\left(x^2-2\sqrt{\alpha^{\ast}}x+|\alpha|^2+
\alpha^{\ast 2}+\frac {2} {q-1}\right)^{
\frac {1} {1-q}}
\otimes\]
\begin{equation}
\label{eq3.22}
\frac {\partial} {\partial x}
\left(x^2-2\sqrt{\alpha}x+|\alpha|^2+
\alpha^2+\frac {2} {q-1}\right)^{
\frac {1} {1-q}}dx
\end{equation}
or
\[<p>_q=-\frac {iA^2(q,\alpha)} {1-q}
\left(\frac {q-1} {2}
\right)^{\frac {2} {1-q}}
\int\limits_{-\infty}^{\infty}
\left(x^2-2\sqrt{\alpha^{\ast}}x+|\alpha|^2+
\alpha^{\ast 2}+\frac {2} {q-1}\right)^{
\frac {1} {1-q}}
\otimes\]
\begin{equation}
\label{eq3.23}
(2x-2\sqrt{2}\alpha)
\left(x^2-2\sqrt{\alpha}x+|\alpha|^2+
\alpha^2+\frac {2} {q-1}\right)^{
\frac {q} {1-q}}dx
\end{equation}
Recourse to (\ref{eqa3}) again, we obtain
\[<p>_q=-\frac {iA^2(q,\alpha)} {1-q}
\left(\frac {q-1} {2}
\right)^{\frac {2} {1-q}}\left[2
\frac {\Gamma\left(\frac {4} {q-1}\right)}
{\Gamma\left(\frac {2q+2} {q-1}\right)}
\right.\otimes\]
\[F_D\left(\frac {4} {q-1}; \frac {1} {q-1},\frac {1} {q-1},
\frac {q} {q-1},
\frac {q} {q-1}; \frac {2q+2} {q-1};1+\beta_1, 1+\beta_2,
1+\beta_3, 1+\beta_4\right)-\]
\[2\sqrt{2}\alpha
\frac {\Gamma\left(\frac {3+q} {q-1}\right)}
{\Gamma\left(\frac {2q+2} {q-1}\right)}\otimes\]
\begin{equation}
\label{eq3.24}
\left. F_D\left(\frac {3+q} {q-1}; \frac {1} {q-1},\frac {1} {q-1},
\frac {q} {q-1},
\frac {q} {q-1}; \frac {2q+2} {q-1};1+\beta_1, 1+\beta_2,
1+\beta_3, 1+\beta_4\right)\right]
\end{equation}

Fig. 1 displays the q-dependence of our four relevant q-mean values.
With the mean q-values  obtained above, we can calculate $(\Delta
x)_q(\Delta p)_q$. The uncertainties are plotted, as a function of $q$, in Fig. 2.

\subsection{$\psi_{\alpha q}$ states form an over-complete basis}

\nd It is easy to see that there is a one-to-one mapping
$|\alpha>\Leftrightarrow|\alpha,q>$, that immediately arises from
the well known one-to-one mapping between q-exponentials and
ordinary ones. This entails that one can write the unity operator
as
\begin{equation}
I=\int\limits_{-\infty}^{\infty}
|\alpha,q>A(q,\alpha)<\alpha,q|\;d^2\alpha= \frac {1} {\pi}
\int\limits_{-\infty}^{\infty} |\alpha><\alpha|\;d^2\alpha,
\end{equation}
with $A(q,\alpha)$ an still unknown constant. Here
$\lim_{q\rightarrow 1}A(q,\alpha)= \frac {1} {\pi}$

\nd Thus, for any q, the basis $\{|\alpha,q>\}$ constitute an over
complete basis.

\setcounter{equation}{0}

\section{Quantum uncertainty in the limit $q \rightarrow 1$}

We will show now that $\lim_{q\rightarrow 1}(\Delta x)_q(\Delta p)_q
=\Delta x\Delta p=\frac {1} {2}$. This is to the essence in order to
ensure that our q-extension of coherent states makes sense.  For
this endeavor we use the approximation, for $q$ close to one, of the
q-exponential. It is easily seen that one has
\begin{equation}
\label{eq3.25} \left[1+i(q-1)z\right]^{\frac {1} {1-q}}=
\left[1-\frac {q-1} {2} z^2\right] e^{-iz}.
\end{equation}
As a consequence of (\ref{eq3.25}) we obtain
\[\left[1+\frac {q-1} {2}(x^2-2\sqrt{2}\alpha x+
\alpha^2+|\alpha|^2)
\right]^{\frac {1} {1-q}}=\]
\[\left[1+\frac {q-1} {8}(x^2-2\sqrt{2}\alpha x+
\alpha^2+|\alpha|^2)^2
\right]
e^{-\frac {1} {2} (x^2-2\sqrt{2}\alpha x+
\alpha^2+|\alpha|^2)}=\]
\begin{equation}
\label{eq3.26}
\left(1+\frac {q-1} {2}\frac {\partial^2} {\partial\beta^2}\right)
e^{-\frac {\beta} {2} (x^2-2\sqrt{2}\alpha x+
\alpha^2+|\alpha|^2)}\left|_{\beta=1}\right.
\end{equation}
The normalized q-coherent state reads, in this approximation,
\begin{equation}
\label{eq3.27} \psi_{\alpha q}(x)=A^{-1}(q,\alpha) \left(1+\frac
{q-1} {2}\frac {\partial^2} {\partial\beta^2}\right) e^{-\frac
{\beta} {2} (x^2-2\sqrt{2}\alpha x+
\alpha^2+|\alpha|^2)}\left|_{\beta=1}\right.
\end{equation}
Of course,  $A(q,\alpha)$  needs evaluation. For this purpose we
calculate
\[A^2(q,\alpha)=
\left(1+\frac {q-1} {2}\frac {\partial^2} {\partial\beta^2}\right)
\left(1+\frac {q-1} {2}\frac {\partial^2} {\partial\gamma^2}\right)
\otimes\]
\begin{equation}
\label{eq3.28}
\int\limits_{-\infty}^{\infty}
e^{-\frac {\beta} {2} (x^2-2\sqrt{2}\alpha^{\ast} x+
\alpha^{\ast 2}+|\alpha|^2)}
e^{-\frac {\gamma} {2} (x^2-2\sqrt{2}\alpha x+
\alpha^2+|\alpha|^2)}
dx\left|_{\beta=\gamma=1}\right.
\end{equation}
By recourse to the Integral-Table result given in \cite{ts1} we
then find
\[A^2(q,\alpha)=\sqrt{\pi}
\left(1+\frac {q-1} {2}\frac {\partial^2} {\partial\beta^2}\right)
\left(1+\frac {q-1} {2}\frac {\partial^2} {\partial\gamma^2}\right)
\otimes\]
\begin{equation}
\label{eq3.29} \left[\frac {\sqrt{2}} {\sqrt{\beta+\gamma}}
e^{-\frac {\beta} {2}(\alpha^{\ast 2}+|\alpha|^2)} e^{-\frac
{\gamma} {2}(\alpha^2+|\alpha|^2)} e^{\frac
{(\beta\alpha^{\ast}+\gamma\alpha)^2}
{\beta+\gamma}}\right]_{\beta=\gamma=1}.
\end{equation}
We can thus write
\begin{equation}
\label{eq3.30} A^2(q,\alpha)=\sqrt{\pi}+(q-1)f_1(\alpha)
+(q-1)^2f_2(\alpha),
\end{equation}
where $f_1$ and $f_2$ are non-singular functions of $\alpha$. As a
consequence,
\begin{equation}
\label{eq3.31} A(q,\alpha)=\sqrt{\sqrt{\pi}+(q-1)f_1(\alpha)
+(q-1)^2f_2(\alpha)},
\end{equation}
and
\begin{equation}
\label{eq3.32} \lim_{q\rightarrow 1}A(q,\alpha)=\pi^{\frac {1}
{4}}.
\end{equation}
We now can write for $<x^2>_q$
\[<x^2>_q=A^{-2}(q,\alpha)
\left(1+\frac {q-1} {2}\frac {\partial^2} {\partial\beta^2}\right)
\left(1+\frac {q-1} {2}\frac {\partial^2} {\partial\gamma^2}\right)
\otimes\]
\begin{equation}
\label{eq3.33}
\int\limits_{-\infty}^{\infty}
e^{-\frac {\beta} {2} (x^2-2\sqrt{2}\alpha^{\ast} x+
\alpha^{\ast 2}+|\alpha|^2)}x^2
e^{-\frac {\gamma} {2} (x^2-2\sqrt{2}\alpha x+
\alpha^2+|\alpha|^2)}
dx\left|_{\beta=\gamma=1}\right.
\end{equation}
Using once more the Integral-Table  \cite{ts1},  one has
\[<x^2>_q=A^{-2}(q,\alpha)
\left(1+\frac {q-1} {2}\frac {\partial^2} {\partial\beta^2}\right)
\left(1+\frac {q-1} {2}\frac {\partial^2} {\partial\gamma^2}\right)
\otimes\]
\begin{equation}
\label{eq3.34} \left\{\frac {2^{\frac {3} {2}}}
{(\beta+\gamma)^{\frac {3} {2}}} e^{-\frac {\beta}
{2}(\alpha^{\ast 2}+|\alpha|^2)} e^{-\frac {\gamma}
{2}(\alpha^2+|\alpha|^2)} e^{\frac
{(\beta\alpha^{\ast}+\gamma\alpha)^2}
{\beta+\gamma}}(2i)^{-2}\sqrt{\pi} H_2\left[\frac
{i(\beta\alpha^{\ast}+\gamma\alpha)}
{\sqrt{\beta+\gamma}}\right]\right\}.
\end{equation}
As a consequence,
\begin{equation}
\label{eq3.35} <x^2>_q=A^{-2}(q,\alpha)\left\{\sqrt{\pi}
\left[\frac {(\alpha+\alpha^{\ast})^2} {2}+ \frac {1}
{2}\right]+(q-1)g_1(\alpha)+ (q-1)^2g_2(\alpha)\right\},
\end{equation}
where $g_1$ and $g_2$ are non-singular functions of $\alpha$.
Thus (See Appendix C),
\begin{equation}
\label{eq3.36} \lim_{q\rightarrow 1} <x^2>_q= \frac {1} {2}+\frac
{(\alpha+\alpha^{\ast})^2} {2}= <x^2>.
\end{equation}
Proceeding now in similar fashion  for $<x>_q$ we obtain
\[<x>_q=A^{-2}(q,\alpha)
\left(1+\frac {q-1} {2}\frac {\partial^2} {\partial\beta^2}\right)
\left(1+\frac {q-1} {2}\frac {\partial^2} {\partial\gamma^2}\right)
\otimes\]
\begin{equation}
\label{eq3.37}
\int\limits_{-\infty}^{\infty}
e^{-\frac {\beta} {2} (x^2-2\sqrt{2}\alpha^{\ast} x+
\alpha^{\ast 2}+|\alpha|^2)}x
e^{-\frac {\gamma} {2} (x^2-2\sqrt{2}\alpha x+
\alpha^2+|\alpha|^2)}
dx\left|_{\beta=\gamma=1}\right.
\end{equation}
According to the Integral-Table result \cite{ts1},
\[<x>_q=A^{-2}(q,\alpha)
\left(1+\frac {q-1} {2}\frac {\partial^2} {\partial\beta^2}\right)
\left(1+\frac {q-1} {2}\frac {\partial^2} {\partial\gamma^2}\right)
\otimes\]
\begin{equation}
\label{eq3.38} \left\{\frac {2} {\beta+\gamma)} e^{-\frac {\beta}
{2}(\alpha^{\ast 2}+|\alpha|^2)} e^{-\frac {\gamma}
{2}(\alpha^2+|\alpha|^2)} e^{\frac
{(\beta\alpha^{\ast}+\gamma\alpha)^2}
{\beta+\gamma}}(2i)^{-1}\sqrt{\pi} H_1\left[\frac
{i(\beta\alpha^{\ast}+\gamma\alpha)}
{\sqrt{\beta+\gamma}}\right]\right\},
\end{equation}
or
\begin{equation}
\label{eq3.39} <x>_q=A^{-2}(q,\alpha)\left\{\sqrt{\pi} \left[\frac
{\alpha+\alpha^{\ast}} {\sqrt{2}} \right]+(q-1)h_1(\alpha)+
(q-1)^2h_2(\alpha)\right\},
\end{equation}
where $h_1$ and $h_2$ are again non-singular functions of
$\alpha$. Accordingly (see Appendix C),
\begin{equation}
\label{eq3.40} \lim_{q\rightarrow 1} <x>_q= \frac
{\alpha+\alpha^{\ast}} {\sqrt{2}}= <x>.
\end{equation}
For $<p^2>_q$ we have instead
\[<p^2>_q=-A^{-2}(q,\alpha)
\left(1+\frac {q-1} {2}\frac {\partial^2} {\partial\beta^2}\right)
\left(1+\frac {q-1} {2}\frac {\partial^2} {\partial\gamma^2}\right)
\otimes\]
\begin{equation}
\label{eq3.41} \int\limits_{-\infty}^{\infty} e^{-\frac {\beta}
{2} (x^2-2\sqrt{2}\alpha^{\ast} x+ \alpha^{\ast 2}+|\alpha|^2)}
\frac {\partial^2} {\partial x^2} e^{-\frac {\gamma} {2}
(x^2-2\sqrt{2}\alpha x+ \alpha^2+|\alpha|^2)}
dx\left|_{\beta=\gamma=1}\right.
\end{equation}
or
\[<p^2>_q=A^{-2}(q,\alpha)
\left(1+\frac {q-1} {2}\frac {\partial^2} {\partial\beta^2}\right)
\left(1+\frac {q-1} {2}\frac {\partial^2} {\partial\gamma^2}\right)
\otimes\]
\begin{equation}
\label{eq3.42}
\int\limits_{-\infty}^{\infty}
e^{-\frac {\beta} {2} (x^2-2\sqrt{2}\alpha^{\ast} x+
\alpha^{\ast 2}+|\alpha|^2)}
[\gamma^2(x-\sqrt{2}\alpha)^2-\gamma]
e^{-\frac {\gamma} {2} (x^2-2\sqrt{2}\alpha x+
\alpha^2+|\alpha|^2)}
dx\left|_{\beta=\gamma=1}\right.
\end{equation}
As in previous cases, according to Integral-Table result
\cite{ts1} we have
\begin{equation}
\label{eq3.43} <p^2>_q=A^{-2}(q,\alpha)\left\{\sqrt{\pi}
\left[\frac {1} {2}- \frac {(\alpha-\alpha^{\ast})^2} {2}
\right]+(q-1)k_1(\alpha)+ (q-1)^2k_2(\alpha)\right\}.
\end{equation}
Here  $k_1$ and $k_2$ are non-singular functions as well.
Therefore (see Appendix C),
\begin{equation}
\label{eq3.44} \lim_{q\rightarrow 1} <p^2>_q= \frac {1} {2}-\frac
{(\alpha-\alpha^{\ast})^2} {2}= <p^2>.
\end{equation}
In analogy with the above case we now also have
\[<p>_q=-iA^{-2}(q,\alpha)
\left(1+\frac {q-1} {2}\frac {\partial^2} {\partial\beta^2}\right)
\left(1+\frac {q-1} {2}\frac {\partial^2} {\partial\gamma^2}\right)
\otimes\]
\begin{equation}
\label{eq3.45} \int\limits_{-\infty}^{\infty} e^{-\frac {\beta}
{2} (x^2-2\sqrt{2}\alpha^{\ast} x+ \alpha^{\ast 2}+|\alpha|^2)}
\frac {\partial} {\partial x} e^{-\frac {\gamma} {2}
(x^2-2\sqrt{2}\alpha x+ \alpha^2+|\alpha|^2)}
dx\left|_{\beta=\gamma=1}\right.
\end{equation}
and, after employing  again the Integral-Table result \cite{ts1},
\begin{equation}
\label{eq3.46} <p>_q=-iA^{-2}(q,\alpha)\left\{\sqrt{\pi}
\left[\frac {\alpha-\alpha^{\ast}} {\sqrt{2}}
\right]+(q-1)l_1(\alpha)+ (q-1)^2l_2(\alpha)\right\},
\end{equation}
where $l_1$ and $l_2$ are non-singular functions of $\alpha$.
Thus (see Appendix C),
\begin{equation}
\label{eq3.47} \lim_{q\rightarrow 1} <p>_q= \frac
{\alpha-\alpha^{\ast}} {i\sqrt{2}}= <p>.
\end{equation}
From (\ref{eq3.36}), (\ref{eq3.40}), (\ref{eq3.44}), and
(\ref{eq3.47}),  we obtain
\begin{equation}
\label{eq3.48} \lim_{q\rightarrow 1} (\Delta x)_q(\Delta p)_q=
\Delta x\Delta p=\frac {1} {2}.
\end{equation}
For the q-distribution, with q close to 1, and using
\begin{equation}
\label{eq3.49} |q,\alpha>=A^{-1}(q,\alpha)\left(1+\frac {q-1} {2}
\frac {\partial^2} {\partial\beta^2}\right)
\int\limits_{-\infty}^{\infty} e^{-\frac {\beta} {2}
(x^2-2\sqrt{2}\alpha^{\ast} x+ \alpha^{\ast 2}+|\alpha|^2)}|x>dx
\left|_{\beta=1}\right.
\end{equation}
we have
\[<k|q,\alpha>=\frac {A^{-1}(q,\alpha)}
{\sqrt{2\pi}}
\left(1+\frac {q-1} {2}
\frac {\partial^2} {\partial\beta^2}\right)\otimes\]
\begin{equation}
\label{eq3.50}
\int\limits_{-\infty}^{\infty}e^{-ikx}
e^{-\frac {\beta} {2} (x^2-2\sqrt{2}\alpha x+
\alpha^2+|\alpha|^2)}dx
\left|_{\beta=1}\right.
\end{equation}
Again, from the Integral-Table result \cite{ts1}, we can write
\begin{equation}
\label{eq3.51} <k|q,\alpha>=A^{-1}(q,\alpha)\left[ e^{-\frac {1}
{2} (k^2+2\sqrt{2}i\alpha k-
\alpha^2+|\alpha|^2)}+(q-1)f(\alpha,k) \right],
\end{equation}
where $f$ is non-singular.
\noindent Using the results given there we have
\begin{equation}
\label{eq3.52} \lim_{q\rightarrow 1}<k|q,\alpha>= \pi^{-\frac {1}
{4}} e^{-\frac {1} {2} (k^2+2\sqrt{2}i\alpha k-
\alpha^2+|\alpha|^2)}=<k|\alpha>,
\end{equation}
and, as a consequence,
\begin{equation}
\label{eq3.53} \lim_{q\rightarrow 1}|<k|q,\alpha>|^2=
|<k|\alpha>|^2= \pi^{-\frac {1} {2}} e^{-(k-p)^2},
\end{equation}
a nice result indeed!

\newpage

\begin{figure}[h]
\begin{center}
\includegraphics[scale=0.6,angle=0]{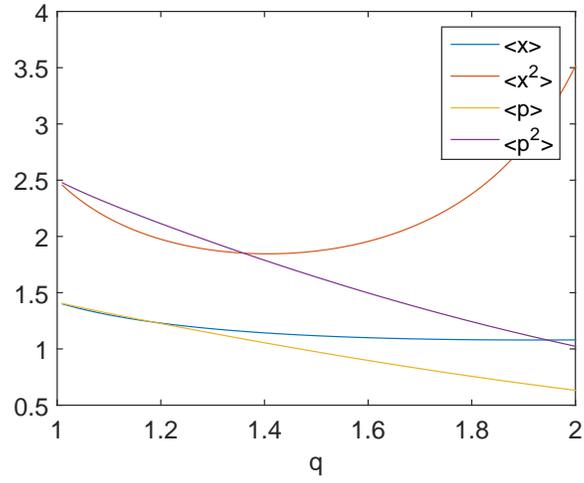}
\vspace{-0.2cm} \caption{Our four relevant mean values are plotted vs. q.}
\end{center}
\end{figure}

\newpage

\begin{figure}[h]
\begin{center}
\includegraphics[scale=0.6,angle=0]{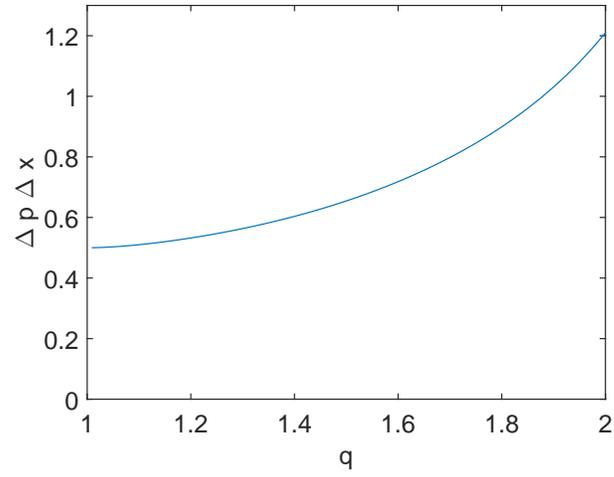}
\vspace{-0.2cm} \caption{Quantum uncertainties vs. q}
\end{center}
\end{figure}

\newpage

\setcounter{equation}{0}

\section{Conclusions}

We have introduced and studied in this work special q-states that
one might denominate Tsallis' pseudo-coherent ones 
(that  {\it have 
 nothing to do with the so-called q-coherent states of 
Quesne, Eremin-Meldianov, and others} \cite{quesne}.)  \vskip 3 mm \nd 
 Also, we obtained some
interesting preliminary results. In particular, we have exhibited
the q-dependence of the quantum uncertainty, that is minimal for
$q=1$.  We emphasize that we have gotten the first overcomplete basis of Tsallis
literature. This should be an interesting addition to such body of work.
 Summing up:

\begin{itemize}

\item  We determined the most important relationships
governing the new Tsallis' pseudo-coherent states.

\item In particular, let us reiterate, we find that, in the limit $q \rightarrow 1 $,
minimal uncertainty is attained (for $q=1$), which constitutes a fundamental
result.

\item We saw that the Tsallis' pseudo-coherent states constitute an
over complete basis for any $q$.

\end{itemize}

\newpage

\renewcommand{\thesection}{\Alph{section}}

\renewcommand{\theequation}{\Alph{section}.\arabic{equation}}

\setcounter{section}{1}
\setcounter{equation} {0}

\section*{Appendix A: Proof of Eq.(\ref{eq2.7})}

It is very well known the annihilation operator for the
one-dimensional harmonic oscillator is given by
\begin{equation}
\label{eqb1}
\hat{a}=\frac {\hat{x}+i\hat{p}} {\sqrt{2}}
\end{equation}
In the x-representation of Quantum Mechanics this operator is
expressed via
\begin{equation}
\label{eqb2}
\hat{a}(x)=\frac {1} {\sqrt{2}}
\left(x+\frac {d} {dx}\right)
\end{equation}
Thus,  a coherent state is defined as the eigenfunction
\begin{equation}
\label{eqb3} \hat{a}(x)\psi_{\alpha}(x)=\frac {1} {\sqrt{2}}
\left(x\psi_{\alpha}(x)+\frac {d\psi_{\alpha}(x)} {dx}\right)=
\alpha \psi_{\alpha}(x),
\end{equation}
or, equivalently,
\begin{equation}
\label{eqb4} \frac {d\psi_{\alpha}(x)}
{dx}=(\sqrt{2}\alpha-x)\psi_{\alpha}(x).
\end{equation}
The solution of (\ref{eqb4}) is
\begin{equation}
\label{eqb5} \psi_{\alpha}(x)=Ce^{-\frac {x^2} {2}}
e^{\sqrt{2}\alpha x}.
\end{equation}
The constant $C$ can be evaluated using the normalization
condition
\begin{equation}
\label{eqb6} \int\limits_{-\infty}^{\infty}|\psi_{\alpha}(x)|^2dx=
|C|^2\int\limits_{-\infty}^{\infty} e^{-x^2}
e^{\sqrt{2}(\alpha+{\alpha}^{\ast}) x}dx=1.
\end{equation}
Accordingly,
\begin{equation}
\label{eqb7} \int\limits_{-\infty}^{\infty}|\psi_{\alpha}(x)|^2dx=
|C|^2e^{\frac {(\alpha+{\alpha}^{\ast})^2} {2}}
\int\limits_{-\infty}^{\infty} e^{-\left(x-\frac
{\alpha+{\alpha}^{\ast}} {\sqrt{2}} \right)^2}dx=1.
\end{equation}
By recourse to the result given in the Table \cite{ts1} we now
obtain
\begin{equation}
\label{eqb8} \int\limits_{-\infty}^{\infty} e^{-\left(x-\frac
{\alpha+{\alpha}^{\ast}} {\sqrt{2}} \right)^2}dx=\sqrt{\pi}.
\end{equation}
As a consequence,
\begin{equation}
\label{eqb9} C=\pi^{-\frac {1} {4}} e^{-\frac
{(\alpha+{\alpha}^{\ast})^2} {4}}.
\end{equation}
Thus,  we have for $\psi_{\alpha}(x)$ the expression
\begin{equation}
\label{eqb10} \psi_{\alpha}(x)=\pi^{-\frac {1} {4}} e^{-\frac
{(\alpha+{\alpha}^{\ast})^2} {4}} e^{-\frac {x^2} {2}}
e^{\sqrt{2}\alpha x},
\end{equation}
or, equivalently,
\begin{equation}
\label{eqb11} \psi_{\alpha}(x)= e^{i\alpha_R\alpha_I}\pi^{-\frac
{1} {4}} e^{-\frac {\alpha^2} {2}} e^{-\frac {|\alpha|^2} {2}}
e^{-\frac {x^2} {2}} e^{\sqrt{2}\alpha x},
\end{equation}
where $\alpha=\alpha_R+i\alpha_I$. As $e^{i\alpha_R\alpha_I}$ is
an imaginary phase, it can be eliminated from (\ref{eqb11}) to
finally obtain
\begin{equation}
\label{eqb12} \psi_{\alpha}(x)=
\pi^{-\frac {1} {4}} e^{-\frac
{\alpha^2} {2}} e^{-\frac {|\alpha|^2} {2}} e^{-\frac {x^2} {2}}
e^{\sqrt{2}\alpha x}.
\end{equation}

\setcounter{section}{2}
\setcounter{equation} {0}

\section*{Appendix B: Lauricella functions}

Lauricella functions $F$ can be regarded as generalizations to 
several variables of the Gauss hypergeometric functions. They were investigated at the end of the 19ts century by Giuseppe Lauricella (1867–1913), an Italian mathematician mostly known by his contribution to elasticity theory.  The
fourth Lauricella function of four variables is given by  \cite{tp5}
\setcounter{equation}{0}
\[F_D(a;b_1,b_2,b_3,b_4;c;x_1,x_2,x_3,x_4)=\]
\[\sum\limits_{m_1=0}^{\infty}
\sum\limits_{m_2=0}^{\infty}
\sum\limits_{m_3=0}^{\infty}
\sum\limits_{m_4=0}^{\infty}
\frac {(a)_{m_1+m_2+m_3+m_4}
(b_1)_{m_1}(b_2)_{m_2}(b_3)_{m_3}(b_4)_{m_4}}
{(c)_{m_1+m_2+m_3+m_4}}\otimes\]
\begin{equation}
\label{eqa1} \frac {x_1^{m_1}x_2^{m_2}x_3^{m_3}x_4^{m_4}} {m_1!
m_2! m_3! m_4!}.
\end{equation}
This function  satisfies \cite{tp5}
\[\int\limits_0^1u^{a-1}(1-u)^{c-a-1}
(1-ux_1)^{-b_1}(1-ux_2)^{-b_2}(1-ux_3)^{-b_3}(1-ux_4)^{-b_4}
du\]
\begin{equation}
\label{eqa2} =\frac {\Gamma(a)\Gamma(c-a)} {\Gamma(c)}
F_D(a;b_1,b_2,b_3,b_4;c;x_1,x_2,x_3,x_4).
\end{equation}
After two variables' changes we can deduce, from  (\ref{eqa2}),
the relation
\[\int\limits_0^{\infty}u^{c-a-1}(1-u)^{b_1+b_2+b_3+b_4-c}
(u_1+z_1)^{-b_1}(u_2+z_2)^{-b_2}(u_3+z_3)^{-b_3}(u_4+z_4)^{-b_4}
du\]
\begin{equation}
\label{eqa3} =\frac {\Gamma(a)\Gamma(c-a)} {\Gamma(c)}
F_D(a;b_1,b_2,b_3,b_4;c;1-z_1,1-z_2,1-z_3,1-z_4).
\end{equation}

\setcounter{section}{3}
\setcounter{equation} {0}

\section*{Appendix C: Reviewing uncertainty relations for coherent states}
For the sake of completeness, we give here some well known results that are
needed in determining uncertainties. For an ordinary  coherent state $|\alpha>$ we have:
\begin{equation}
\label{eqc1}
<x^2>=\pi^{-\frac {1} {2}}
\int\limits_{-\infty}^{\infty}
e^{-\frac {1} {2} (x^2-2\sqrt{2}\alpha^{\ast} x+
\alpha^{\ast 2}+|\alpha|^2)}x^2
e^{-\frac {1} {2} (x^2-2\sqrt{2}\alpha x+
\alpha^2+|\alpha|^2)}dx
\end{equation}
With the use of the Integral-Table result \cite{ts1} we find
\begin{equation}
\label{eqc2}
<x^2>=(2i)^{-2}
H_2\left[\frac {i(\alpha^{\ast}+\alpha)}
{\sqrt{2}}\right]
\end{equation}
and then
\begin{equation}
\label{eqc3}
<x^2>=
\frac {1} {2}+\frac {(\alpha+\alpha^{\ast})^2} {2}
\end{equation}
For $<x>$ the situation is quite similar
\begin{equation}
\label{eqc4}
<x>=\pi^{-\frac {1} {2}}
\int\limits_{-\infty}^{\infty}
e^{-\frac {1} {2} (x^2-2\sqrt{2}\alpha^{\ast} x+
\alpha^{\ast 2}+|\alpha|^2)}x
e^{-\frac {1} {2} (x^2-2\sqrt{2}\alpha x+
\alpha^2+|\alpha|^2)}dx
\end{equation}
Using the Integral-Table result \cite{ts1} again we obtain
\begin{equation}
\label{eqc5}
<x>=(2i)^{-1}
H_1\left[\frac {i(\alpha^{\ast}+\alpha)}
{\sqrt{2}}\right]
\end{equation}
and thus
\begin{equation}
\label{eqc6}
<x>=
\frac {\alpha+\alpha^{\ast}} {\sqrt{2}}
\end{equation}
For $<p>$, the integral is somewhat more complicated
\begin{equation}
\label{eqc7}
<p^2>=-\pi^{-\frac {1} {2}}
\int\limits_{-\infty}^{\infty}
e^{-\frac {1} {2} (x^2-2\sqrt{2}\alpha^{\ast} x+
\alpha^{\ast 2}+|\alpha|^2)}
\frac {\partial^2} {\partial x^2}
e^{-\frac {1} {2} (x^2-2\sqrt{2}\alpha x+
\alpha^2+|\alpha|^2)}dx
\end{equation}
or:
\begin{equation}
\label{eqc8}
<p^2>=\pi^{-\frac {1} {2}}
\int\limits_{-\infty}^{\infty}
e^{-\frac {1} {2} (x^2-2\sqrt{2}\alpha^{\ast} x+
\alpha^{\ast 2}+|\alpha|^2)}
[1-(x-\sqrt{2}\alpha)^2]
e^{-\frac {1} {2} (x^2-2\sqrt{2}\alpha x+
\alpha^2+|\alpha|^2)}dx
\end{equation}
Now,  by recourse to the Integral-Table result \cite{ts1} we
obtain
\begin{equation}
\label{eqc9}
<p^2>=1-2\alpha^2-i\sqrt{2}\alpha
H_1\left[\frac {i(\alpha^{\ast}+\alpha)}
{\sqrt{2}}\right]+\frac {1} {4}
H_2\left[\frac {i(\alpha^{\ast}+\alpha)}
{\sqrt{2}}\right]
\end{equation}
or
\begin{equation}
\label{eqc10}
<p^2>=
\frac {1} {2}-\frac {(\alpha-\alpha^{\ast})^2} {2}
\end{equation}
For dealing with $<p>$ one starts with
\begin{equation}
\label{eqc11}
<p>=-i\pi^{-\frac {1} {2}}
\int\limits_{-\infty}^{\infty}
e^{-\frac {1} {2} (x^2-2\sqrt{2}\alpha^{\ast} x+
\alpha^{\ast 2}+|\alpha|^2)}
\frac {\partial} {\partial x}
e^{-\frac {1} {2} (x^2-2\sqrt{2}\alpha x+
\alpha^2+|\alpha|^2)}dx
\end{equation}
or
\begin{equation}
\label{eqc12}
<p>=i\pi^{-\frac {1} {2}}
\int\limits_{-\infty}^{\infty}
e^{-\frac {1} {2} (x^2-2\sqrt{2}\alpha^{\ast} x+
\alpha^{\ast 2}+|\alpha|^2)}
(x-\sqrt{2}\alpha)
e^{-\frac {1} {2} (x^2-2\sqrt{2}\alpha x+
\alpha^2+|\alpha|^2)}dx
\end{equation}
and, finally,
\begin{equation}
\label{eqc13}
<p>=\frac {\alpha-\alpha^{\ast}} {i\sqrt{2}}
\end{equation}
Accordingly,  the well-known uncertainty relation for a coherent
state becomes
\begin{equation}
\label{eqc14} \Delta x\Delta p=\frac {1} {2},
\end{equation}
i.e., minimal uncertainty, the main feature of coherent state.

\end{document}